\documentstyle[editedvolume,psfig]{crckapb10} 

\font\smfont=cmr7
\font\smtfont=cmr5
\font\smttfont=cmr5

\newfam\smfam

\textfont\smfam=\smfont
\scriptfont\smfam=\smtfont
\scriptscriptfont\smfam=\smttfont

\def\spose#1{\hbox to 0pt{#1\hss}}
\def\arcsec{\ifmmode {^{\prime\prime ~}}\else $^{\prime\prime ~}$\fi}
\def\arcmin{\ifmmode {^{\prime}}\else $^{\prime}$\fi}
\def\arcsper{\ifmmode \rlap.{^{\prime\prime}}\else
    $\rlap.{^{\prime\prime}}$\fi}
\def\arcmper{\ifmmode \rlap.{^{\prime}}\else
    $\rlap.{^{\prime}}$\fi}

\def\deg{\ifmmode {^{\circ}}\else {$^\circ$}\fi}
\def\degr{\ifmmode {^{\circ}}\else {$^\circ$}\fi}
\def\degs{\ifmmode {^{\circ}}\else {$^\circ$}\fi}

\def\eg{{\it e.g.} }
\def\ergps{\ifmmode {\rm\,erg\,s^{-1}}\else ${\rm\,erg\,s^{-1}}$\fi}
\def\ergpscm2{\ifmmode {\rm\,erg\,s^{-1}\,cm^{-2}}\else
    ${\rm\,erg\,s^{-1}\,cm^{-2}}$\fi}
\def\ergpscm2A{\rm\,erg\,s^{-1}\,cm^{-2},{\AA}^{-1}}
\def\etal{{\it et al.~}}

\def\h3Mpc{h^{-3}{\rm Mpc}^3}

\def\H0{{\rm\,H$_\circ$~}}

\def\kms{\ifmmode {\rm\,km\,s^{-1}}\else
    ${\rm\,km\,s^{-1}}$\fi}

\def\Lya{{\rm\,Ly-$\alpha$~}}

\def\minper{\ifmmode \rlap.{^{m}}\else $\rlap{.}{^m} $\fi}

\def\q0{{\rm\,q$_\circ$~}}

\def\secper{\ifmmode \rlap.{^{s}}\else $\rlap{.}{^{s}} $\fi}

\def\spose#1{\hbox to 0pt{#1\hss}}
\def\simlt{\mathrel{\spose{\lower 3pt\hbox{$\mathchar"218$}}
     \raise 2.0pt\hbox{$\mathchar"13C$}}}
\def\simgt{\mathrel{\spose{\lower 3pt\hbox{$\mathchar"218$}}
     \raise 2.0pt\hbox{$\mathchar"13E$}}}

\def\araa{{ARA\&A}}

\def\aasup{{A\&AS}}
\def\aj{{AJ}}
\def\apj{{ApJ}}

\def\apjs{{ApJS}}

\def\mnras{{MNRAS}}

\def\pasp{{PASP}}

\def\apjref#1;#2;#3;#4 {\par\pp#1, {#2}, #3, #4 \par}

\begin{opening}
\title{Searches for High Redshift Radio Galaxies}

\author{Carlos De Breuck}
\author{Wil van Breugel}
\institute{Institute for Geophysics and Planetary Physics,
Lawrence Livermore National Laboratory, L$-$413, P.O. Box 808,
Livermore, CA 94550, U.S.A.}
\author{Huub R\"ottgering}
\author{George Miley}
\institute{Leiden Observatory, P.O. Box 9513, 2300 RA Leiden, The Netherlands}

\end{opening}

\runningtitle{USS High$-$Redshift Radio Galaxy Searches}

\begin{document}

\begin{abstract}

We have started a search for High Redshift Radio Galaxies (HZRGs) in
an area covering 7 sr by selecting a sample of Ultra Steep Spectrum
(USS) sources with a low flux density cut-off S$_{1400} >$ 10 mJy
and a steep spectral index cut-off of $\alpha < -1.3$ ($S \propto
\nu^{\alpha}$) from the WENSS, NVSS and TEXAS surveys. Our first
results for 27 sources show that we are almost twice as effective in
finding HZRGs than than surveys of relatively bright radio sources
with a spectral index cut-off of $\alpha < -1.0$. The redshift
distribution is consistent with an extension of the $z - \alpha$
relation to $\alpha < -1.3$, but a large fraction of our sample (40\%)
consists of objects which are too faint to observe with 3$-$4m class
telescopes. Our search is aimed at increasing the number of very high
redshift radio galaxies for further detailed studies of the formation
and evolution of massive galaxies and their environment.

\end{abstract} 

\section{Introduction}
High Redshift Radio Galaxies (HZRGs) have significantly contributed to
our understanding of the formation of galaxies and their environments
\cite{mcc93}.  Unlike quasars, the light from HZRGs is not dominated by
the (non-stellar) AGN component but is spatially resolved, allowing us
to get a clearer view of the host galaxy and its young stellar
population, using HST and Keck.  A dramatic example of this is the
recent discovery of jet-induced star formation in the HZRG 4C41.17 at
$z=3.8$ \cite{dey97,wvb97a}, and the discovery of fully formed HZRG
ellipticals at $z=2.5$ \cite{wvb97b}.

Unfortunately, the number of known HZRGs dramatically decreases at high
redshift (Figure 1).  At present, 120 Radio Galaxies are known with $z
 >$ 2, of which 16 have $z >$ 3 and only 3 have $z >$ 4.  Determining
redshifts of large numbers of radio sources is very demanding in
telescope time.  Samples of optically faint radio sources therefore
have to be limited, thereby unavoidably introducing selection effects.

To obtain significant samples of HZRGs, two main strategies have been
used in the past: (i) surveys of a restricted part of the sky using a
low flux density cut-off; (ii) all-sky surveys with a radio spectral
index or radio-`colour' bias.  The first major search for HZRGs was
the 3CR survey, which now has complete redshift information
\cite{spi85,raw96a}.  This survey is biased as it only includes
sources of extreme radio power (S$_{151} >$ 10 Jy), imposing a
tight redshift-power relation, which makes it difficult to disentangle
redshift and power dependence (\eg Best, Rawlings, these proceedings).
A southern equivalent to this 3CR survey, but 10 times deeper, is
the Molonglo Reference Catalog/1 Jy survey \cite{mcc96}.  In finding
very high redshift radio galaxies, the MRC/1Jy is not efficient, as $<
1 \%$ of the galaxies are at $z >$ 3.  A similar survey, the 6C/B2 is
being undertaken in the northern hemisphere (Rawlings, these
proceedings; Eales \etal, 1997)

More effective searches for HZRGs have been made using with samples of
Ultra Steep Spectrum (USS; $\alpha < -1$) radio sources covering a
large sky area.  Several samples of USS sources \cite{cha96,rot94} have
led to the discovery of more than 70 $z>2$ radio galaxies, more than
half of the 120 known today.  Figure 1 shows the contribution of USS
selected sources (shaded) to the total number of $z>2$ HZRGs.  All the
highest redshift radio galaxies ($z > 3.5$) have been found using USS
criteria.

\begin{figure}
\centerline{
\psfig{figure=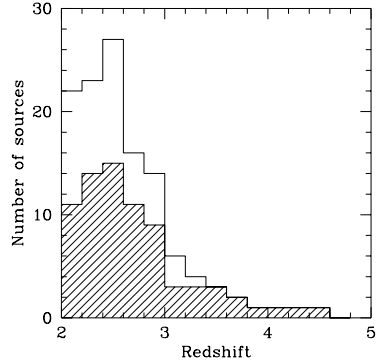,width=5cm}
}
\caption{\label{zgt2fig} 
Known $z>2$ Radio Galaxies. USS  sources are shaded.} 
\end{figure}

\section{The sample of USS sources from the WENSS, NVSS and FIRST}

With the advent of the WENSS, NVSS and FIRST surveys it is now possible for
the first time to define a uniformly selected sample of USS sources
covering the entire sky North of declination $-$40\deg, and using 10 -- 100 
times lower flux density limits than has been possible before.

\subsection{The WENSS/NVSS (WN) sample}
A correlation of 1997 March versions of the WENSS and NVSS catalogues
provided spectral indices for 78,000 sources, of which we selected the
1\% with spectral indices $\alpha^{325}_{1400} < -1.3$.  Several
additional cut-offs were applied to the sample: (i) S$_{1400} > 10$
mJy, to obtain a complete sample with spectral indices accurate to
0.1; (ii) galactic latitude $b >$ 15\degr, to avoid excessive galactic
extinction during optical imaging and spectroscopy; and (iii) only
unresolved sources in the WENSS survey (resolution 54\arcsec) were
retained.  Unlike some other USS samples (\eg the 6C$^*$; Rawlings,
these proceedings), we did not apply other size cut-offs, as HZRGs
exist with angular sizes of 30\arcsec or more (our sample contains for
example a $z=3.215$ galaxy of 35\arcsec; Bremer, these proceedings).
The present WN sample consists of 284 USS sources.

\subsection{The TEXAS/NVSS (TN) sample}
The WENSS survey covers the northern quarter of the sky.  In order to
extend our samples to the southern hemisphere, while still maintaining
uniform sample, we correlated the TEXAS 365 MHz survey \cite{dou96}
with the NVSS. Next to the same cut-offs as in the WN sample, we have
also excluded sources of questionable fluxes or positions from the
TEXAS catalogue \cite{dou96}.  Because we are including TEXAS sources
down to the flux limit, the TN sample will only be $\sim 80\%$
complete at S$_{365}$ = 250 mJy. As a result, the TN sample goes 4
times less deep than the WN sample. In the area covered by both
WENSS and TEXAS, we therefore preferred the WENSS fluxes. Our TN
sample covers $-40$\degr $< \delta <$ 28\degr. We estimated the errors
in the TN spectral indices by comparing them with the WN spectral
indices in the overlapping region.  We found that there is no
systematic difference between TN and WN spectral indices as a function
of WENSS flux density.

\subsection{High-Resolution Radio Images from FIRST and VLA observations}
Positional uncertainties of NVSS sources become larger as flux
decreases (Condon, these proceedings).  As we are reaching flux levels
where these errors become $\sim 3$\arcsec, and virtually all our
objects are fainter than $m_R = 20$, the NVSS position accuracy is
inadequate for optical identifications.  We therefore correlated our
WN sample with the FIRST catalogue \cite{bec95} which has accurate
positions, and provides a check of the 1.4 GHz flux density, allowing
us to weed out variable sources.  Unfortunately, the present coverage
of the FIRST survey overlaps with only 25\% of our WN/TN samples.  We
therefore observed 239 sources in our sample outside the FIRST area
with the VLA in A and BnA arrays. Observations were done at 4.8 GHZ to
increase the resolution relative to the FIRST survey and to identify
spectral curvature.

\subsection{Optical Spectroscopy}
The first year of spectroscopic observations of our sample using 3 $-$ 4
m class telescopes has yielded firm redshifts for 27 sources. The
median redshift is $z=2.38$; only 8 have $z < 2$, while 15 have $2 < z
< 3$, 3 have $3 < z < 4$, and 1 recently discovered has $z=4.11$. For
almost 40\% of the sources observed, we could not yet determine the
redshift because either no object or no emission lines were detected.
We will attempt to obtain redshifts of these extremely high redshift
candidates using larger aperture telescopes. An indication that these
objects are indeed that distant is the K $>$ 20.5 limit we found for 2
of them, suggesting redshifts significantly beyond $z=4$ on the basis of
the K$-z$ relation.

\section{Discussion}
\subsection{Redshift $-$ Spectral Index Relation}
Figure 2 plots spectral index against redshift for 4 different samples
of radio sources. The 3CR and MRC surveys (dots and crosses) have no
spectral index bias, but still show the spectral index $-$ redshift
correlation. This relation is due to a combination of k$-$correction and
intrinsic spectral curvature of the radio spectra (Carilli {\it et al}, in
preparation).  The 4C sample (open circles; $\alpha_{178}^{1415}
< -1$; median $z=1.84$) shows a rather uniform redshift distribution of
sources, from $z=0.4$ to 3.8, which includes a dramatically higher
percentage of HZRGs than the surveys without spectral index bias.  The
initial results of our WN/TN sample have shown that we are even more
efficient in finding $z > 2$ sources than the 4C (Table 1). The major
differences between our sample and the 4C are our steeper spectral
index cut-off and our lower flux limit.  The higher median redshift in
our sample can be interpreted as a combination of two effects: (i) an
extension of the redshift $-$ spectral index relation towards steeper
spectral$-$indices; and (ii) our lower flux density limit causing the
objects in our S$_{1400} >$ 10 mJy flux$-$limited sample to be more
distant than the S$_{178} >$ 2 Jy 4C sample.

\begin{figure}
\centerline{
\psfig{file=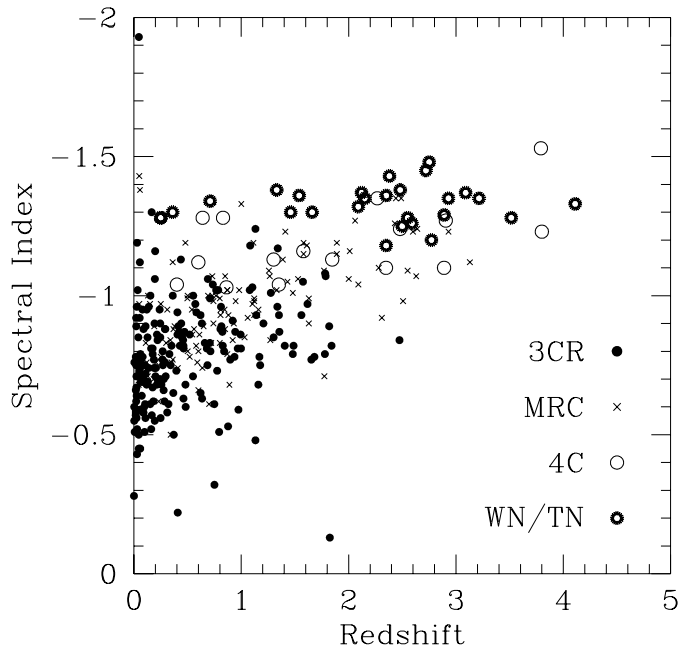,width=9cm}
}
\caption{\label{zafig} 
$\alpha_{1400}^{325}$ against $z$ for 4 samples.} 
\end{figure}

\begin{table}[htb]
\begin{center}
\caption{Efficiencies in finding HZRGs}
\begin{tabular}{ c  c  c  c  c }\hline \hline
 & \multicolumn{3}{c}{Known Redshift} & Unknown \\ 
\hline
 & $z<2$ & $2<z<3$ & $z>3$ &  \\
\hline
 3CR    & 99.5 \% & 0.5 \% &   0 \% &  0 \% \\
 MRC    & 89.4 \% & 16  \% & 0.6 \% & 40 \% \\
 B2/6C  & 87.7 \% & 8.8 \% & 3.5 \% & 10 \% \\
 4C     & 53   \% &  35 \% &  12 \% & 50 \% \\
 WN/TN  & 30   \% &  55 \% &  15 \% & 40 \% (faint) \\
\hline
\end{tabular}
\end{center}
\end{table}

\subsection{Radio Powers}
The advantages of using radio galaxies instead of quasars to study the
formation of giant ellipticals are their much large angular sizes and
much fainter non-stellar AGN.  However, also radio galaxies have a
residual contribution of non-stellar light from the central AGN,
confusing our view of the stellar population (Rawlings, these
proceedings). A direct indication of this dependence is the correlation
between emission-line luminosity and radio power \cite{mcc93}. Our
initial results also show a rather weak correlation between the \Lya
luminosity and radio power.  The key to minimizing the effects of radio
AGN activity is therefore to probe lower powers, which we achieve with
the much lower flux density limit of our sample.  Figure 3 shows that
we are indeed beginning to find galaxies at these lower powers.  The
new galaxies at $z > 2$ with P$_{325} \simlt 10^{28}$ W/Hz/sr will
allow us to disentangle the redshift and power evolution of radio
galaxies, with the caveat of using primarily USS sources at high
redshift.

\begin{figure}
\centerline{
\psfig{file=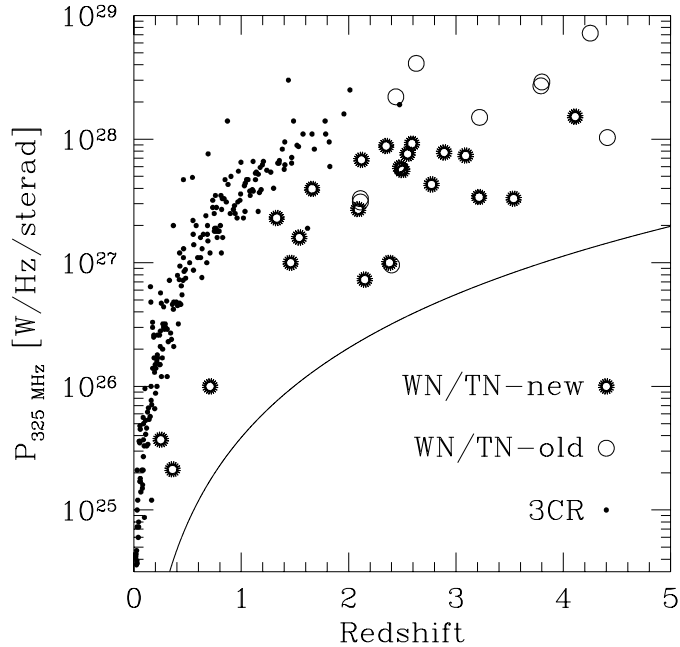,width=9cm}
}
\caption{\label{zpfig} 
P$_{325}$ against $z$ for 3CR and our sample. Old WN/TN sources are
known $z >$ 2 radio galaxies falling into our sample. The line indicates the lower limit of our WN sample.} 
\end{figure}

\section{Conclusions}

The new radio surveys have allowed us to define a sample of faint
(S$_{1400} >$ 10 mJy) sources with $\alpha < -1.3$.  We have found
more $z > 2$ sources than previous USS or other surveys, indicating
the $z - \alpha$ relation extends out to higher redshifts. The
fraction of $z < 2$ galaxies is less than 1/3, indicating we are near
the peak efficiency of the USS technique. Stellar age estimates of the
parent ellipticals of HZRGs all indicate formation redshifts $z_F >$
5.  The 40\% of our sources that have remained undetected with 3$-$4m
class telescopes are prime targets for finding these.

\begin{acknowledgements}
Work performed at the Lawrence Livermore National Laboratory is
supported by the DOE under contract W7405-ENG-48
\end{acknowledgements}


\begin{thebibliography}{} 

\bibitem[\protect\citeauthoryear{Becker {\it et al.}}{1995}]{bec95} 
Becker, R., White, R., Helfand, D., 1995,  {\apj}\/ 450, 559. 

\bibitem[\protect\citeauthoryear{Chambers {\it et al.}}{1996}]{cha96} 
Chambers, K., Miley, G., van Breugel, W., \& Huang, J., 1996,  {\apjs}\/ 106, 215. 

\bibitem[\protect\citeauthoryear{Dey {\it et al.}}{1997}]{dey97} 
Dey, A. \etal 1997, in preparation

\bibitem[\protect\citeauthoryear{Douglas {\it et al.}}{1996}]{dou96} 
Douglas, J., Bask, F., Bozyan, F., Torrence, G., \& Wolfe, C., 1996, {\aj}\/ 111, 1945.

\bibitem[\protect\citeauthoryear{Eales \& Rawlings}{1997}]{eal97}
Eales, S., Rawlings, S., Law-Green, D., Cotter, G., \& Lacy, M., 1997, {\mnras}, astro-ph 9701023.

\bibitem[\protect\citeauthoryear{McCarthy {\it et al.}}{1993}]{mcc93}
McCarthy, P., 1993, {\araa}\/ 31, 639.

\bibitem[\protect\citeauthoryear{McCarthy {\it et al.}}{1996}]{mcc96}
McCarthy, P., Kapahi, V., van Breugel, W., Persson, S., Athreya, R., \& Subrahmanya, C., 1996, {\apjs}\/ 107, 19.

\bibitem[\protect\citeauthoryear{Rawlings {\it et al.}}{1996a}]{raw96a}
Rawlings, S., Lacy, M. Leahy, J., Dunlop, J., Garrington, S. \& Ludke, E., 1996, {\mnras}\/ 279, 13.

\bibitem[\protect\citeauthoryear{R\"ottgering {\it et al.}}{1994}]{rot94}
R\"ottgering, H., Lacy, M., Miley, G., Chambers, K., Saunders, R., 1994, {\aasup}\/ 108, 79.

\bibitem[\protect\citeauthoryear{Spinrad {\it et al.}}{1985}]{spi85}
Spinrad, H., Djorgovski, S., Marr, J, \& Aguilar, L., 1985, {\pasp}\/ 97, 932.

\bibitem[\protect\citeauthoryear{van Breugel {\it et al.}}{1997}]{wvb97a} 
van Breugel, W. \etal, 1997, in preparation

\bibitem[\protect\citeauthoryear{van Breugel {\it et al.}}{1997}]{wvb97b}
van Breugel, W. \etal, 1997, in preparation

\end{thebibliography}
\end{document}